\documentclass[preprint,12pt]{aastex}

\shortauthors{Holman et al.\ 2006}
\shorttitle{Transit Photometry of XO-1b}

\begin{document}

%
\newcommand{\C}{\ensuremath{^{\circ}C\;}}
\newcommand{\el}{\ensuremath{e^-}}
\newcommand{\sqarcsec}{\ensuremath{\Box^{\prime\prime}}}
\newcommand{\sqarcdeg}{\ensuremath{\Box^{\circ}}}
\newcommand{\aduel}{\ensuremath{\lbrack ADU/\el \rbrack}}
\newcommand{\eladu}{\ensuremath{\rm {\lbrack \el/ADU \rbrack}}}
\newcommand{\adupixs}{\ensuremath{\rm ADU/(pix\, s)}}
                                                                                          
%

\bibliographystyle{apj}

\title{
The Transit Light Curve (TLC) Project.\\
I.~Four Consecutive Transits of the Exoplanet XO-1b
}

\author{
Matthew J.\ Holman\altaffilmark{1},
Joshua N.\ Winn\altaffilmark{2},
David W.\ Latham\altaffilmark{1},\\
Francis T.\ O'Donovan\altaffilmark{3},
David Charbonneau\altaffilmark{1,7},
G\'asp\'ar A.~Bakos\altaffilmark{1,8},\\
Gilbert A.\ Esquerdo\altaffilmark{1,4},
Carl Hergenrother\altaffilmark{5,1},
Mark E.\ Everett\altaffilmark{4},
Andr\'as P\'al\altaffilmark{6,1}
}

\altaffiltext{1}{Harvard-Smithsonian Center for Astrophysics, 60
  Garden Street, Cambridge, MA 02138; mholman@cfa.harvard.edu}
\altaffiltext{2}{Department of Physics, and Kavli Institute for
  Astrophysics and Space Research, Massachusetts Institute of
  Technology, Cambridge, MA 02139}
\altaffiltext{3}{California Institute of Technology, 1200 East
  California Blvd., Pasadena, CA 91125}
\altaffiltext{4}{Planetary Science Institute, 1700 East Fort Lowell,
  Tucson, AZ 85719}
\altaffiltext{5}{Lunar and Planetary Laboratory, University of
  Arizona, Tucson, AZ 85719}
\altaffiltext{6}{E\"otv\"os Lor\'and University}
\altaffiltext{7}{Alfred P.\ Sloan Research Fellow.}
\altaffiltext{8}{Hubble Fellow.}

\begin{abstract}

  We present $RIz$ photometry of four consecutive transits of
  the newly discovered exoplanet XO-1b.  We improve upon the estimates of
  the transit parameters, finding
  the planetary radius to be
  $R_{\rm P} = 1.184_{-0.018}^{+0.028}~R_{\rm Jup}$,
  and the stellar radius to be
  $R_{\rm S} = 0.928_{-0.013}^{+0.018}~R_\odot$,
  assuming a stellar mass of
  $M_{\rm S}=1.00 \pm 0.03~M_\odot$. The
  uncertainties in the planetary and stellar radii are dominated
  by the uncertainty in the stellar mass.
  These uncertainties increase by a factor of 2--3
  if a more conservative uncertainty
  of $0.10~M_\odot$ is assumed for the stellar mass.  Our estimate of the
  planetary radius is smaller than that reported by \citet{McCullough.2006} and yields a mean density that is 
  comparable to that of TrES-1 and HD~189733b.  The timings of the transits have an accuracy ranging
  from 0.2 to 2.5~minutes, and are marginally consistent with a uniform period.

\end{abstract}

\keywords{planetary systems --- stars:~individual (GSC~02041-01657) ---
techniques: photometric}

\section{Introduction}

An exoplanetary transit is a rare opportunity to learn a great deal
about both the planet and the star.  With precise measurements of the
amount of light blocked by the planet as a function of time, it is
possible to infer the relative sizes of the star and planet, the
orbital inclination, and the stellar limb-darkening function. Coupled
with measurements of the time-variable Doppler shift of the star and
an estimate of the stellar mass, one learns the planetary mass and the
stellar radius. These fundamental measurements set the stage for a
host of more subtle measurements of effects such as planetary
atmospheric absorption lines, thermal emission, spin-orbit alignment,
and timing anomalies, as reviewed recently by
\citet{Charbonneau.2006c}.

For these reasons, newly-discovered transiting exoplanets are welcomed
with open arms. The tenth such object was recently reported by
\citet{McCullough.2006}. The parent star, XO-1, is bright ($V=11$,
G1~V), making it a favorable target for precise observations. The
planet has an orbital period of $\sim$4~days and a mass and radius
comparable to Jupiter, although \citet{McCullough.2006} point out that
their photometry actually implies a mean density that is somewhat
smaller than theoretical expectations for ``hot Jupiters.'' If
confirmed, this would put XO-1b in the same category as the
anomalously large planet HD~209458b, and may have implications for the
various theories that have been espoused for that object.

Two of us (M.J.H.\ and J.N.W.) have initiated the Transit Light Curve
(TLC) Project, a long-term campaign to build a library of
high-precision transit photometry, with the dual goals of (1) refining
the estimates of the physical and orbital 
parameters of the target systems, and (2) searching for secular and
short-term variations in the transit times (and light curves) that
would be indicative of perturbations from additional
bodies~\citep{Agol.2005,Holman.2005a}. Here, we present results for
XO-1b that were obtained as part of this program. We describe the
observations and the data reduction procedures in \S~2. In \S~3 we
describe the model and techniques we used to estimate the physical and
orbital parameters of the XO-1 system, and in \S~4 we summarize our
results.

\section{The Observations and Data Reduction}

We observed four consecutive transits of XO-1b.  According to the
ephemeris provided by McCullough et al.~(2006),
\begin{equation}
T_c(E) = 2,453,808.9170~\mathrm{[HJD]} + E\times (3.941534~\mathrm{days}),
\end{equation}
these transits correspond to epochs 17 through 20. We employed three different
telescopes: the FLWO~1.2m telescope (for $E=19,20$); the
Palomar~1.5m telescope (for $E=17$), and the TopHAT~0.26m telescope
(for $E=17,18,19$).

\begin{boldmath}
\subsection{FLWO~1.2m $z$ Photometry}
\end{boldmath}

We observed the $E=19,20$ transits (UT~2006~May~28 and June~1) with
KeplerCam on the 1.2m (48~inch) telescope of the Fred L.\ Whipple
Observatory (FLWO) on Mt.~Hopkins, Arizona. This camera
(P.I.\ D.~Latham) was built for a photometric survey of the target field of the {\it Kepler}\/ satellite
mission~\citep{Borucki.2003}. It has a single $4\mathrm{K}\times4\mathrm{K}$ Fairchild 486
CCD with a $23\farcm 1 \times 23\farcm 1$ field of view.  We used $2
\times 2$ binning, for which the readout and reset time is 11.5~s and the typical read noise
is 7~$e^{-}$ per binned pixel.  The response of each amplifier
deviates from linearity by less that 0.5\% over the range of counts from
the faintest to brightest comparison star.  We observed through the SDSS $z$
filter, the reddest available band, in order to minimize the effects
of limb-darkening and color-dependent atmospheric extinction. The effective bandpass at the red end was limited
by the quantum efficiency of the CCD, which drops from $\sim 100$\% at
$7500~\AA$ to $\sim 10$\% at $10500~\AA$.  We defocused the telescope
slightly in order to enhance the duty cycle and average over
pixel-to-pixel sensitivity variations. The full-width at half-maximum
(FWHM) of a stellar image was typically $\sim$3 binned pixels
($2\arcsec$). We used automatic guiding to maintain the locations of
XO-1 and its comparison stars to within a few pixels over the course
of both nights. On each night, we repeatedly took 30~second exposures
for approximately 5~hr bracketing the predicted transit midpoint.
The conditions on UT~2006~May~28 appeared photometric, and
the images were taken through airmasses ranging from 1.00 to 1.22. The conditions on UT~2006~June~1
were nearly photometric, except for very thin, high clouds that passed
through the field between UT~05:50 and UT~06:30. The airmass range
on this night
was from 1.00 to 1.31.

The images were calibrated using standard IRAF\footnote{IRAF is
  distributed by the National Optical Astronomy Observatories, which
  are operated by the Association of Universities for Research in
  Astronomy, Inc., under cooperative agreement with the National
  Science Foundation.} procedures for the overscan correction,
trimming, bias subtraction, and flat-field division.  We did not
attempt to correct the fringing that was apparent with the $z$ filter.
The fringing had a small amplitude and little effect on the final
photometry, given the accuracy of the automatic guiding.  We excluded
three images the first night that showed significant large-scale
features that were not corrected by the flat-field (presumably from
clouds), and one from the second night that was contaminated by an
artificial satellite trail.  We then performed aperture photometry of
XO-1 and 4 nearby stars of comparable brightness and color (stars 1,
3, 6, and 7 from Table~1 of \citealt{McCullough.2006}), using an
aperture radius of 6.5 pixels ($4\farcs 3$) for both nights. We
subtracted the underlying contribution from the sky, after estimating
its brightness within an annulus ranging from 30 to 35 pixels in
radius, centered on each star.  We divided the flux of XO-1 by the
total flux of the comparison stars.  We then fit a linear function of
time to the pre-ingress and post-egress data, and divided the entire
time series by this function, in order to normalize the flux and
remove residual systematic effects. A function of time proved to be a
slightly better fit than the more traditional function of airmass.

To estimate the uncertainties in our photometry, we computed the
quadrature sum of the errors due to Poisson noise of the stars (both
XO-1 and the comparison stars), Poisson noise of the sky background,
readout noise, and scintillation noise (as estimated according to the
empirical formulas of~\citealt{Young.1967} and
\citealt{Dravins.1998}). The dominant term is the Poisson noise from
XO-1.  The final time series is plotted in Fig.~1 and is available in
electronic form in Table~1.

\clearpage

\begin{figure}[p]
\includegraphics[scale=1, angle=90.]{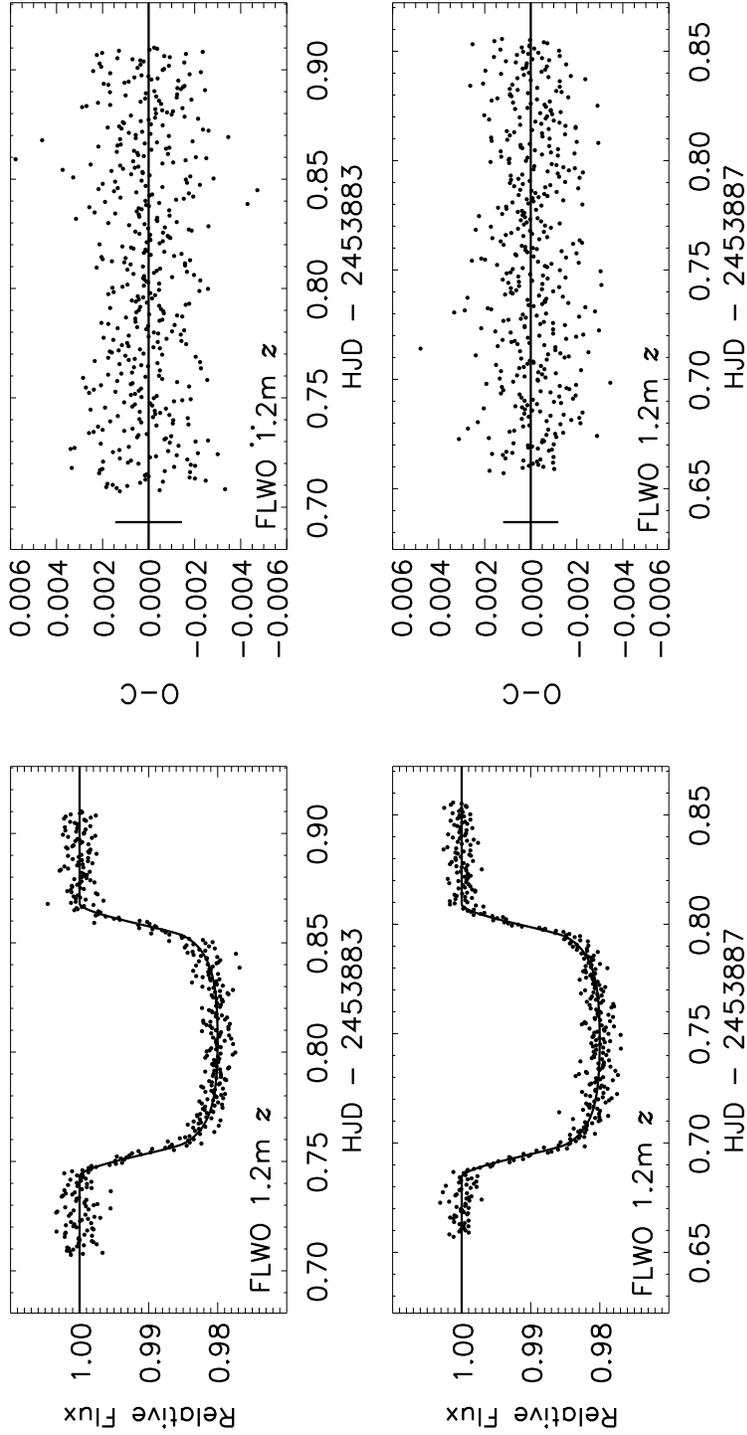}
\caption{Relative $z$ band photometry of XO-1. The left panels show the
photometry (points) and the best-fitting model (solid line). The right
panels show the residuals (observed~$-$~calculated) and 
representative $1~\sigma$ error bars that have been rescaled such that
the $\chi^2$ per degree of freedom for each time series is unity.  
For the first transit ($E=19$), the RMS residual is 0.15\%.
For the second transit, the RMS residual is 0.12\%.
\label{fig:lc1}}
\end{figure}

\clearpage

\begin{boldmath}
\subsection{TopHAT $I$ Photometry}
\end{boldmath}

We used TopHAT to observe the $E=17,18,19$ transits of XO-1b
(UT~2006~May~20, 24, and 28). TopHAT is an automated telescope, also
located on Mt.~Hopkins, Arizona, which was designed for photometric
follow-up observations of transiting exoplanet candidates identified
by the HAT Network~\citep{Bakos.2004}.  It consists of a 0.26m
diameter f/5 commercially-available Baker Ritchey\--Chr\'etien
telescope on an equatorial fork mount. A 1$\fdg$25-square field of
view is imaged onto a 2K$\times$2K Peltier-cooled, thinned CCD
detector, yielding a pixel scale of 2$\farcs$2. In order to extend the
integration times and increase the duty cycle of the observations, we
applied a slight defocusing. 
The resulting PSF had a FWHM of
2.1 pixels (4$\farcs$6). On each night, we observed for approximately
5~hrs.

We calibrated the images by subtracting the overscan bias and a scaled
dark image, and dividing by an average sky flat from which large
outliers had been rejected. 
We performed aperture photometry on XO-1 and on an additional $\sim$800
stars in the field, using an aperture of radius 5~pixels (11$\farcs$0),
and an exterior annulus for sky subtraction. Most of the 800 stars
(after removing variables) were used as calibrators in a
statistically-weighted manner to transform the magnitudes of the
individual frames to the instrumental magnitude system of a selected
reference frame. The derived light-curve still suffers from
small-amplitude systematic errors.  In order to minimize them, we 
used all the out-of-transit data (assuming it to be of constant
brightness) to find the correlation with the airmass, hour-angle (linear fits), the pixel
position (sinusoidal function), and the Gaussian profile parameters
(second order fits). The fitted function was then applied to and 
subtracted from the entire light-curve, including the transits.  These
corrections in this post-processing step were of the order of 3~mmag or less
for unsaturated points. 
The resulting time series is shown in Fig.~2 and listed in Table~1,
along with the Palomar data described below. 

\clearpage

\begin{figure}[p]
\epsscale{1.0} 
\plotone{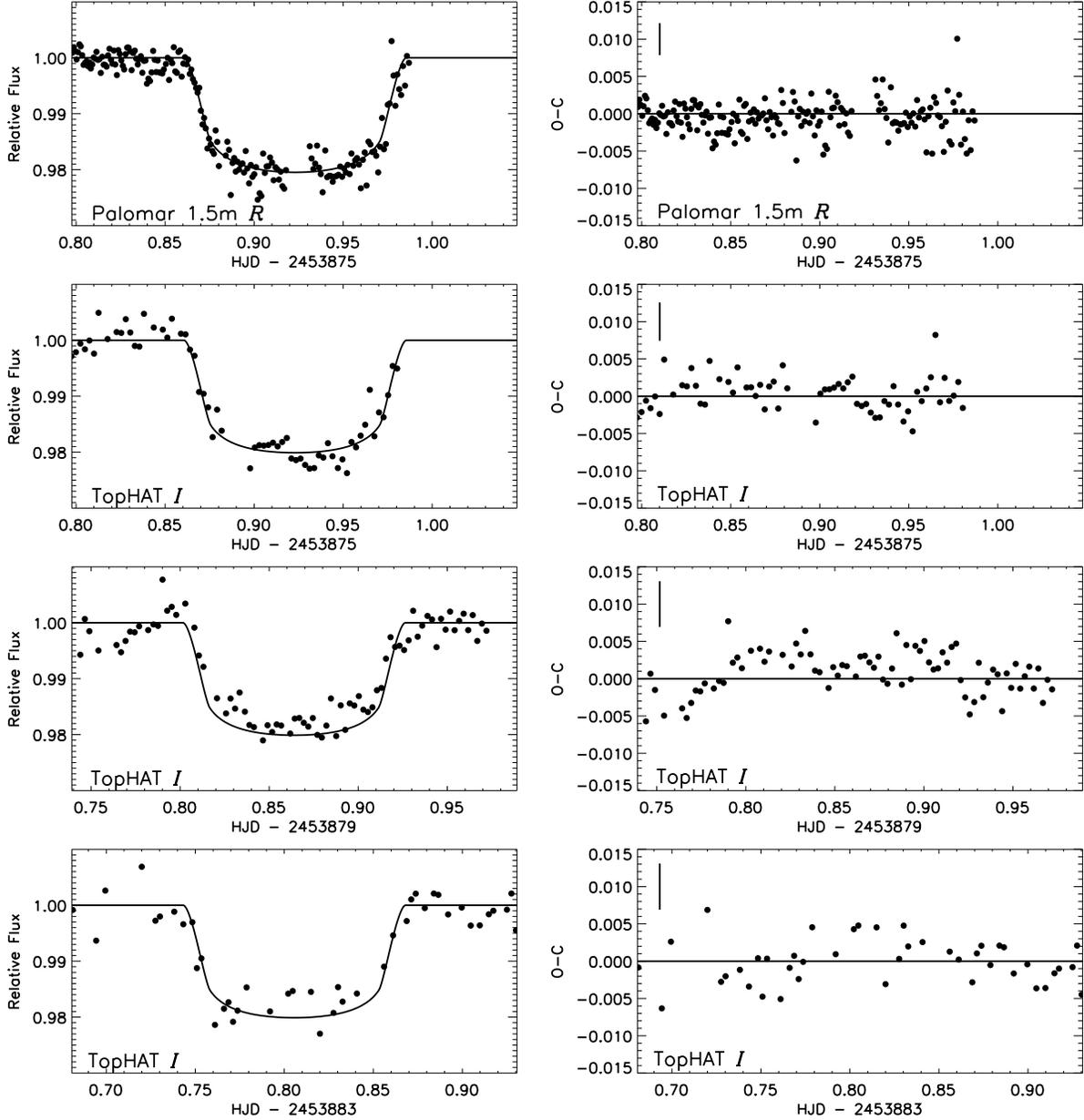}
\caption{Relative $R$ and $I$ band photometry of XO-1. The left panels show the
photometry (points) and the best-fitting model (solid line). The right
panels show the residuals (observed~$-$~calculated) and 
representative $1~\sigma$ error bars, estimated as described in the
text. From top to bottom, the RMS residuals
are 0.21\%, 0.28\%, 0.31\%, and 0.32\%.
\label{fig:lc2}}
\end{figure}

\clearpage

\begin{boldmath}
\subsection{Palomar~1.5m $R$ Photometry}
\end{boldmath}

We observed the $E=17$ transit with the 1.5m (60~inch) telescope at
Palomar Observatory. The CCD camera has 2K$\times$2K~pixels with a
plate scale of $0\farcs$378~pixel$^{-1}$. In order to increase the
duty cycle, we read out only half of the available field of view,
yielding an effective field size of $12{\farcm}9 \, \times \,
6{\farcm}5$. The sky was cloud-free, the typical seeing was
$1\farcs5$, and the observations ranged in airmass from 1.01$-$1.45.
We gathered 209 $R$-band images spanning 5.2~h, with integration times
of 20~s and a cadence that increased from 65~s to 90~s over the course
of the observing sequence. The observing sequence was briefly
interrupted twice (once prior to ingress, and once near mid-transit)
by telescope calibration scripts that are required as part of its
robotic operation.

We used the automated P60 reduction pipeline to calibrate the images.
This pipeline trims the overscan columns, subtracts the bias level,
divides by a dome flat, and flags bad pixels. We performed aperture
photometry of XO-1 and 5 comparison stars, using an aperture of radius
9~pixels ($3{\farcs}4$), and an annulus for sky subtraction ranging in
radius from 30 to 45 pixels. We divided the flux of XO-1 by the sum of
fluxes of the comparison stars, and normalized the resulting time
series to produce a flux of unity in the pre-ingress data. The RMS
variation of the pre-ingress data is 0.17\%, which we adopted as the
photometric uncertainty in each data point.  The pointing drifted by
as much as 35 pixels ($13{\arcsec}$) during the course of the
observations.  We believe that this drift, coupled with uncertainties
in the flat-field image, is responsible for the poorer quality of the
Palomar~1.5m data compared to those from the FLWO~1.2m. The Palomar
data are presented in Fig.~2 and listed in Table~1.

\section{The Model}

The planetary, stellar, and orbital parameters were inferred by
fitting a parameterized model to all of the photometry
simultaneously. The model is based on a star and a planet on a
circular orbit about their center of mass.\footnote{We assume a
  circular orbit because the radial velocity data show no evidence for
  a non-zero eccentricity, and in the absence of any evidence for
  additional bodies in the system, it is expected that tidal
  interactions have had sufficient time to circularize the orbit (see,
  e.g., \citealt{Rasio.1996,Trilling.2000,Dobbs-Dixon.2004}).} The
star has a mass $M_{\rm S}$ and radius $R_{\rm S}$, and the planet has
a mass $M_{\rm P}$ and radius $R_{\rm P}$. The orbit has a period $P$
and an inclination $i$ relative to the sky plane.  We define the
coordinate system such that $0\arcdeg \leq i\leq 90\arcdeg$. The
initial condition is specified by $T_c$, a particular time of
conjunction (the transit midpoint). When the planet is projected in
front of the star, the model flux declines by an amount that depends
on the projected separation, on the stellar limb-darkening function,
and also on the planet-to-star area ratio. To compute this flux
decrement, we assume a quadratic limb-darkening law, and employ the
analytic formulas of \citet{Mandel.2002}.  The $R$ and $I$ band data
are not of sufficiently high signal-to-noise ratio to justify this
treatment; instead, we fix both combinations of the $R$ and $I$
limb-darkening parameters at the values estimated by
\citet{Claret.2004} for a star of the observed effective temperature,
surface gravity, and metallicity (see Table~5 of
\citealt{McCullough.2006}).

We allow the $T_c$ for each of the 4 transits to be an independent
parameter. This is because we seek to measure or bound any timing
anomalies that may indicate the presence of moons or additional
planets in the system. Obviously if we allow each of the four $T_c$
values to vary, we cannot independently determine the orbital
period. Instead, we fix $P=3.941534$~days, the value reported by
\citet{McCullough.2006}. This mean period is based on observations
spanning a few years and is known more accurately than we could hope
to determine from our time baseline of 12 days. The quoted uncertainty
in the mean period is only 0.000027~days and is negligible for our
purposes.

There is a well-known degeneracy among $M_{\rm S}$, $R_{\rm S}$ and
$R_{\rm P}$ that prevents all three parameters from being uniquely
determined from transit photometry alone, unless a stellar mass-radius
relation is assumed~\citep{Seager.2003}. We fix $M_{\rm
  S}=1.0~M_\odot$, based on the spectroscopic estimate by
\citet{McCullough.2006}. Our results may be scaled to other choices
for the stellar mass according to $R_{\rm S} \propto (M_{\rm
  S}/M_\odot)^{1/3}$ and $R_{\rm P} \propto (M_{\rm
  S}/M_\odot)^{1/3}$. The planetary mass $M_{\rm P}$ is irrelevant to
the model except for its minuscule effect on the relation between $P$
and the semimajor axis; for completeness, we assume $M_{\rm
  P}=0.9~M_{\rm Jup}$, again following \citet{McCullough.2006}.

In total, there are 8 free parameters describing 1309 photometric data
points. The free parameters are $R_{\rm S}$, $R_{\rm P}$, and $i$; the
$z$-band limb darkening parameters $u_1$ and $u_2$; and the four
values of $T_c$.  In practice, we found it better to fit for the
parameters $2u_1 + u_2$ and $u_1 - 2u_2$, because the resulting
uncertainties in those parameters are uncorrelated (as will be shown
below).  We allowed the limb darkening parameters to range only over
the values that produce a monotonically decreasing intensity from the
center of the star to the limb.

Prior to fitting the full set of observations, we fitted each of the
six time series separately and determined the minimum $\chi^2$ in each
case.  The resulting values of $\chi^2$ per degree of freedom were:
2.03 (FLWO~1.2m night 1), 1.39 (FLWO~1.2m night 2), 1.53
(Palomar~1.5m), 0.25 (TopHAT night 1), 0.50 (TopHAT night 2), and 0.34
(TopHAT night 3). Thus, it seems that the calculated uncertainties
were somewhat underestimated for the FLWO and Palomar data, and
overestimated for the TopHAT data.  Before proceeding, we scaled the
estimated uncertainties of each time series individually so that the
resulting value of $\chi^2$ per degree of freedom was unity. Table~2
gives the uncertainties {\it after} this scaling was performed.

We determined the best-fitting model using the AMOEBA
algorithm~\citep{Press.1992} to minimize the error statistic
\begin{equation}
\chi^2 = \sum_{j=1}^{1309}
\left(
\frac{f_j({\mathrm{obs}}) - f_j({\mathrm{calc}})}{\sigma_j}
\right)^2,
\end{equation}
where $f_j$(obs) is the flux observed at time $t_j$, $\sigma_j$ is the
corresponding uncertainty, and $f_j$(calc) is the calculated value.
In Figs.~1 and 2, the left-hand panels show the best-fitting model as
a solid line, and the right-hand panels show the results of
subtracting the calculated values from the observed values. The FLWO
$z$ band data show random-looking residuals with a standard deviation
of 0.15\% for the first night, and 0.12\% for the second night. Almost
all of the leverage on the stellar and planetary parameters comes from
these data. The P60 data also show nearly random residuals, but at a
higher level of 0.2\%, and with occasional outliers. The TopHAT data
are noisier, with a standard deviation of 0.3\%, and show some signs
of correlated residuals (i.e.\ systematic errors). The uncertainties
in the fitted parameters were estimated using two different methods,
described below.

The first method was a bootstrap analysis, similar to those we have
performed previously for the transiting exoplanets
HD~209458b~\citep{Winn.2005b}, OGLE-TR-10~\citep{Holman.2005b}, and
HD~149026b~\citep{Charbonneau.2006a}. We refitted the parameters to
each of $10^4$ different ``realizations'' of the data. These
realizations were sets of 1309 data points drawn randomly from the
actual data set, with duplications allowed (i.e., with
replacement). Each realization was required to preserve the total
number of points in each of the 6 individual time series. The
resulting collection of $10^4$ optimized parameter sets was taken to
be the joint probability distribution for the parameters.

The second method was a Markov Chain Monte Carlo (MCMC) simulation. In
this method (described lucidly for astrophysicists by
\citealt{Tegmark.2004} and \citealt{Ford.2005}), a stochastic process
is used to create a sequence of points in parameter space that
approximates the desired probability distribution. The sequence, or
``chain,'' is generated from an initial point by iterating a ``jump
function.'' In our case the jump function was the addition of a
Gaussian random number to each parameter value. If the new point has a
lower $\chi^2$ than the previous point, the jump is executed; if not,
the jump is only executed with probability
$\exp(-\Delta\chi^2/2)$. Under fairly benign mathematical assumptions,
the chain will eventually converge to the desired probability
distribution. To speed convergence, the Gaussian perturbations should
be large, but not so large that all jumps are rejected. We set the
relative sizes of the perturbations using the 1~$\sigma$ uncertainties
estimated previously by the bootstrap method, and we set the overall
jump size by requiring that $\sim$25\% of jumps are executed. We
created 10 independent chains, each from a random initial position
$\sim$5$\sigma$ away from the optimized parameter values. Each chain
had 500,000 points, the first 20\% of which were discarded to minimize
the effect of the initial condition. The typical correlation length
for each parameter (see~\citealt{Tegmark.2004}) was $\sim$400 points,
giving an effective length of $\sim$1000 per chain.  To check the
convergence and the consistency between chains, we computed the
\citet{Gelman.1992} statistic for each parameter, which is a
comparison between the inter-chain variance and the intra-chain
variance. The results were within a few per cent of unity, a sign of
good mixing and convergence.

The two methods produced very similar results. We also checked the
results for the planetary and stellar radii using the more traditional
approach of stepping each parameter through a sequence of values while
allowing all of the other parameters to float, and identifying the
values for which $\Delta\chi^2 = 1$ as the 68\% confidence limits (as
done by Brown et al.~2001, among others).  Again, the values and the
uncertainties were comparable to the results of the MCMC and bootstrap
methods.  For brevity, we report only the MCMC results for the
remainder of this paper.  The probability distributions for some of
the parameters are shown in Fig.~3, and some of the correlations
between the parameters are shown in Fig.~4. Table~2 lists the median
value of each parameter with their 68\% confidence limits, based on
the MCMC results.

\clearpage

\begin{figure}[p]
\epsscale{0.8}
\plotone{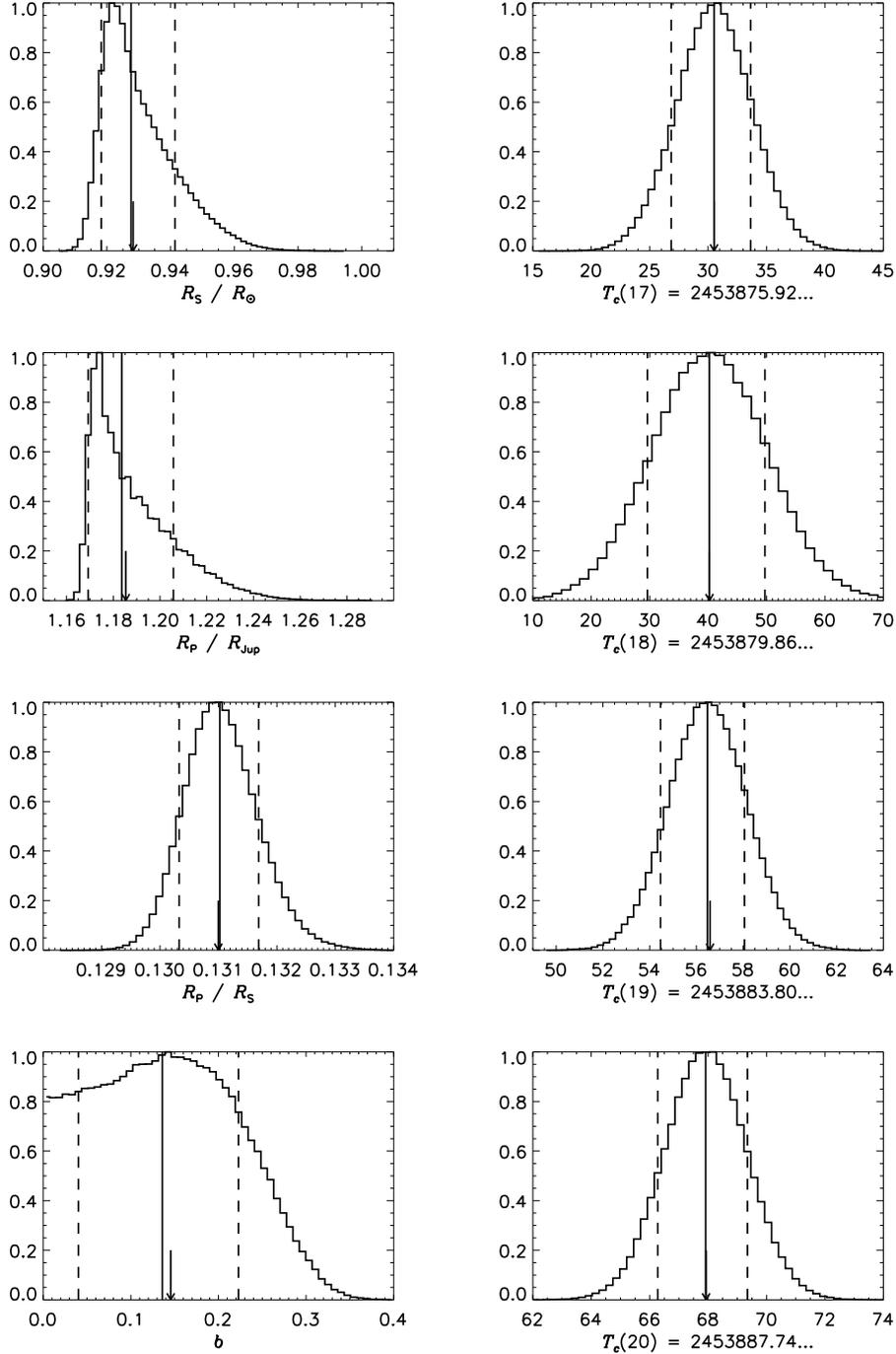}
\caption{ Estimated probability distributions of some planetary,
stellar, and orbital parameters. The histograms show the
results of 10 Markov Chain Monte Carlo simulations,
each with 400,000 points. The median of each distribution is indicated with a
solid line. The dashed lines enclose
68\% of the results, with equal probability on either side of the the median. The arrows show the
the choice of parameters that minimizes $\chi^2$.
The numbers in Table~2 are the median values,
with confidence limits given by the dashed lines.
\label{fig:probdist}}
\end{figure}

\begin{figure}[p]
\epsscale{1.0}
\plotone{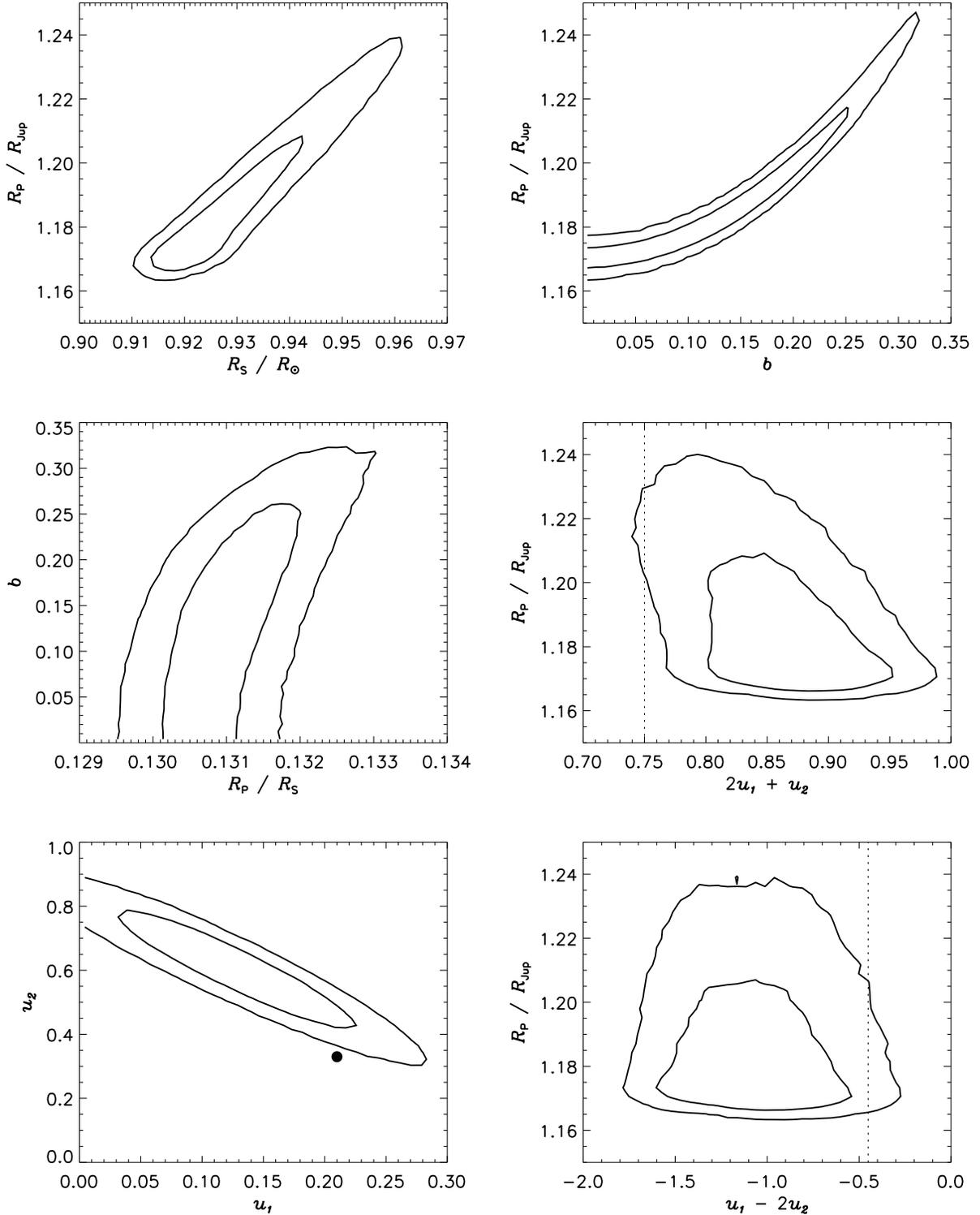}
\caption{ Joint probability distributions of some planetary, stellar,
orbital, and limb-darkening parameters, based on the MCMC simulations.
The contours are isoprobability contours enclosing
68\% and 95\% of the points in the Markov chains.
In the $u_1$--$u_2$ plot, the dot shows the values calculated
by \citet{Claret.2004}.  In the other plots involving limb
darkening, the dotted lines shows the values calculated
by \citet{Claret.2004}.
\label{fig:correlations}}
\end{figure} 

\clearpage

\section{Summary and Discussion}

Through observations of four consecutive transits, we have
significantly improved upon the estimates of the system parameters of
XO-1. The most interesting parameters are the radius of the star, the
radius of the planet, and the mid-transit times, which will be
discussed shortly. The results for the other parameters are not
especially interesting but they do seem reasonable. The results for
the orbital inclination are best described as bounds on the impact
parameter $b$, which is the minimum projected star-planet distance, in
units of the stellar radius. It is given by $b=a \cos i / R_{\rm S}$,
where $a$ is the orbital distance. The data favor a central transit,
with $b< 0.27$ and $i<88\fdg53$ at the 95\% confidence level.

Although the survey and follow-up photometry of
\citet{McCullough.2006} were impressive, and built a convincing case
for an exoplanet, those authors did not attempt to fit for the stellar
radius when modeling the transit light curve.  Instead, they used the
value $R_{\rm S}/R_\odot=1.00\pm 0.08$, based on an interpretation of
the stellar spectrum.  This is because the inference of $R_{\rm S}$
from a transit light curve requires that the ingress and egress are
well sampled and measured with a high signal-to-noise ratio. This type
of data was not available.  The higher precision and finer time
sampling of our data, and of the $z$ band data in particular, allow
for the determination of $R_{\rm S}$ from the light curve, without
relying on spectral modeling and theoretical isochrones.  The
resulting ``photometric'' value of $R_{\rm S}$ is still subject to a
systematic error due to the covariance with $M_{\rm S}$, but the
dependence is fairly weak, $R_{\rm S} \propto M_{\rm S}^{1/3}$,
generally leading to a smaller uncertainty in $R_{\rm S}$ than can be
achieved from spectral modeling and theoretical isochrones.

Our result is $R_{\rm S}/R_\odot = 0.928_{-0.013}^{+0.018}$, which is
consistent with (but more precise than) the value determined by
\citet{McCullough.2006}.  Here we have incorporated the $0.03~M_\odot$
uncertainty in $M_{\rm S}$ determined by \citet{McCullough.2006}.  We
note that this radius is somewhat small for the G1~V spectral type of
XO-1, but it is still consistent, given the stated uncertainties.  We
remind the reader again that the quoted result assumes $M_{\rm
  S}=1.0~M_\odot$, and that the inferred $R_{\rm S}$ scales as
$(M_{\rm S}/M_\odot)^{1/3}$.  From the Yonsei-Yale isochrones, a
stellar mass of $M_{\rm S}= 0.96~M_\odot$ corresponds to a radius of
$R_{\rm S}=0.91~R_\odot$, for solar metallicity and an arbitrary age
of 3.6~Gyr, ~\citep{Yi.2001}.  Thus, a $1.3~\sigma$ change in the
estimated stellar mass yields an estimated stellar radius that is
precisely in line with theoretical expectations.  We also note that
the stellar radius uncertainty is a factor of 2--3 larger if a more
conservative uncertainty of $0.10~M_\odot$ is assumed for the stellar
mass, as shown in Table~2.

Our derived radius of XO-1b is $R_{\rm P}/R_{\rm Jup} =
1.184_{-0.018}^{+0.028}$ (again assuming the uncertainty in the stellar
mass to be $0.03~M_\odot$).  Previously, \citet{McCullough.2006} found
$R_{\rm P}/R_{\rm Jup} =1.30\pm 0.11$. These figures are also in agreement
right within their respective 68\% confidence
limits. Interestingly, we obtain very precise agreement with
\citet{McCullough.2006} for all parameters if we first time-average
our data into 5-minute bins (i.e., by a factor of 8, for the $z$
band data). The \citet{McCullough.2006} data were averaged into bins
ranging in width from 3 to 9 minutes, depending on the telescope
used. We suggest that it is possible that some of the previous
results were slightly biased by the coarser time sampling of the
photometry.  Resolving the degeneracy among the stellar radius,
planetary radius, and orbital inclination requires adequate sampling
of ingress and egress.  

The uncertainties in the limb darkening parameters $u_1$ and
$u_2$ are highly correlated, with the linear combination $2u_1 + u_2$
being well constrained by the data, and the orthogonal combination
$u_1 - 2u_2$ being weakly constrained by the data (see the lower
left panel of Fig.~3).
We find
$2 u_1 + u_2 = 0.86 \pm 0.05$.  This is $2~\sigma$ larger
than the value based on the theoretical
calculations of \citet{Claret.2004},
which predict
$u_1 = 0.21$,
$u_2 = 0.33$, and $2u_1 + u_2 = 0.75$
(for the standard $z$ band, $T=5750~\rm K$, $\log g = 4.5$, $[M/H] = 0.05$, and microturbulent
velocity $v_t = 2.0~\mathrm{km/s}$).
The theoretical values are shown in Fig.~3 as the solid
symbol in the $u_1$--$u_2$ plot, and as dotted lines
in the other two limb-darkening plots.
One might consider
reducing the number of degrees
of freedom and adopting the \citet{Claret.2004} values
as fixed quantities.
When we do so,
we find
$R_{\rm S}/R_\odot = 0.94$,
$R_{\rm P}/R_{\rm Jup} = 1.22$, and $b = 0.26$ ($i = 88.65$~deg), with
the minimum $\chi^2$ increased by 9.
However, given the quality of
the $z$ band data, the unknown level of uncertainty in the theoretical
values, and the possible differences between the FLWO48/KeplerCam
$z$ band and the standard SDSS $z$ band,
we believe fitting for the limb-darkening coefficients is
more appropriate.

Some of the probability distributions shown in Fig.~3 are
asymmetric. This is typical of all fits to transit light curve
data. The fact that the orbital inclination has a maximum value
(namely, 90\arcdeg), combined with the measured durations of the
ingress, egress, and the full transit, imposes this asymmetry among
the covariant parameters $R_{\rm S}$, $R_{\rm P}$ and $b$.

Our downward revision of the planetary radius translates into
an increased value for the mean density,
$ 0.67 \pm 0.07  \ {\rm g \, cm^{-3}}$.  This value is $45$--$56$\%
that of Jupiter. This is comparable
to, but slightly less than, the mean densities of TrES-1
($0.84 \ {\rm g \, cm^{-3}}$; \citealt{Sozzetti.2004}) and
HD~189733b ($0.93 \ {\rm g \, cm^{-3}}$; \citealt{Bakos.2006}).
For XO-1b's estimated equilibrium temperature $T_{\rm eq} = 1100~{\rm K}$
(assuming Bond albedo $A_B = 0.4$ and our derived value of the stellar
radius $R_{\rm S} = 0.928~R_\odot$) and its mass $M_{\rm P} =
0.9~M_J$, the models of \citet{Bodenheimer.2003} predict planetary radii of $R_{\rm P}
= 1.04~R_J$ and $1.11~R_J$ for models with and without a 20~$M_\oplus$
core, respectively.  Our estimate of the radius of XO-1b is
$2~\sigma$ larger than its predicted value, even for a planet without a
core.  HD~189733b's measured radius $R_{\rm P} = 1.154
\pm 0.032~R_J$ \citep{Bakos.2006} is also larger than its theoretical value
($R_{\rm P} = 1.03~R_J$ with a core, $R_{\rm P} = 1.11~R_J$ without a core), given its equilibrium temperature  $T_{\rm
  eq} = 1050~{\rm K}$ and mass $M_{\rm P} = 0.82 \pm
0.03$~\citep{Bouchy.2005}.   In contrast, \citet{Laughlin.2005} showed
that TrES-1's measured radius $R_{\rm P} = 1.08 \pm 0.05~R_J$
\citep{Laughlin.2005} is consistent with its theoretically predicted
value ($R_{\rm P} = 1.05~R_J$ with a core, $R_{\rm P} = 1.09~R_J$
without a core).   

The measured radii of both XO-1b and HD~189733 are consistent with the
predictions of \citet{Bodenheimer.2003} if ``kinetic heating'' is
included.  In these models $\sim 2$\% of the stellar insolation is
deposited at depth, following the work of \citet{Guillot.2002}.  
The age of XO-1 is also uncertain; the planet would be larger if the
system were younger~\citep{Burrows.2003}.
Whether or not the radius of XO-1b requires an additional
energy source, as is the case for HD~209458b
($0.36 \ {\rm g \, cm^{-3}}$; \citealt{Knutson.2006}),
is an important topic for future theoretical work.
The kinetic heating model, proposed to explain the apparent
inflation of HD~209458b, would naturally predict that many other ``hot
Jupiters'' should be inflated.  Other explanations, such as ongoing tidal
circularization due to an eccentricity exchange with a third body
\citep{Bodenheimer.2001}, or the trapping in a Cassini state with
nonzero obliquity \citep{Winn.2005a}, would seemingly have difficulty
accounting for a large population of inflated objects.  

The accuracy of our transit times ranges from 0.2~minutes for the FLWO
$z$ band observations to 2.5~minutes for the transit
observed solely by TopHAT\@.  Fig.~5 shows the differences between the observed
and predicted times of mid-transit, as a function of transit epoch. The predicted times 
assume the average orbital period determined by
\citet{McCullough.2006} and a reference time based on our
observations. So far, all the times are marginally consistent with a
constant period.  These observations provide accurate anchors for
future searches for transit time variations. 

\clearpage

\begin{figure}[p]
\epsscale{1.0} 
\plotone{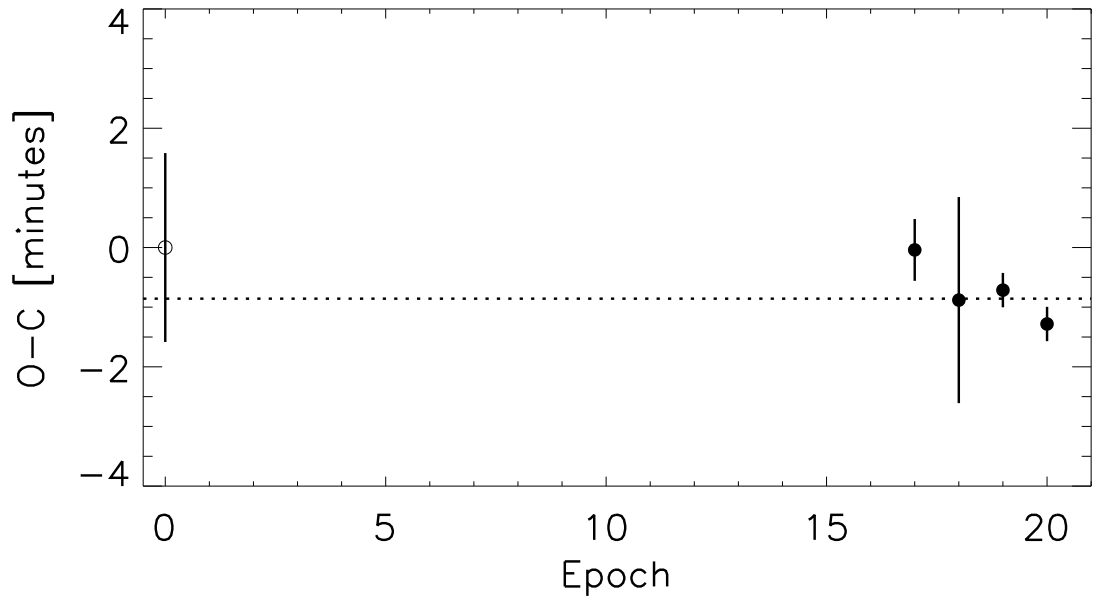}
\caption{ The timing residuals for the 4 observed transits,
according to the ephemeris of \citet{McCullough.2006}.
(see Eq.~1). The first point corresponds to the $T_c$ of
\citet{McCullough.2006}.  The points lie on a horizontal line,
and therefore the data are marginally consistent with a constant period.
\label{fig:mass_radius}}
\end{figure} 

\clearpage

\acknowledgments We thank T.\ Spahr for swapping telescope nights on
short notice, E.\ Falco for accommodating our observing schedule
changes, S.\ Gaudi for suggesting the MCMC method, J. Fern\'andez for
helpful discussions about fitting transit light curves, G.\ Torres for
help with stellar isochrones, and R.\ Kurucz for calculations of
limb-darkening coefficients. We also thank the anonymous referee for
an exceptionally careful review of the manuscript. KeplerCam was
developed with partial support from the Kepler Mission under NASA
Cooperative Agreement NCC2-1390 (P.I.\ D.~Latham), and the KeplerCam
observations described in this paper were partly supported by grants
from the Kepler Mission to SAO and PSI\@. The TopHAT observations were
supported by NASA grant NNG04GN74G. Work by G.B.\ was supported by
NASA through grant HST-HF-01170.01-A, awarded by the Space Telescope
Science Institute, which is operated by the Association of
Universities for Research in Astronomy, Inc., for NASA, under contract
NAS 5-26555.  Work by F.T.O'D. and D.C.\ was supported by NASA under
grant NNG05GJ29G, issued through the Origins of Solar Systems Program.


\clearpage

\begin{deluxetable}{lcccc}
\tabletypesize{\normalsize}
\tablecaption{Photometry of XO-1\label{tbl:photometry}}
\tablewidth{0pt}

\tablehead{
\colhead{Telescope} & \colhead{Filter} & \colhead{HJD} & \colhead{Relative flux} & \colhead{Uncertainty} \\
}

\startdata
         FLWO48 & $z$ & $2453883.70727$   &     $1.00125$   &        $0.00147$   \\
      Palomar60 & $R$ & $2453875.77023$   &     $1.00144$   &        $0.00215$   \\
         TopHAT & $I$ & $2453875.64414$   &     $1.00889$   &        $0.00386$   \\
\enddata 

\tablecomments{The time stamps represent the Heliocentric Julian Date
  at the time of mid-exposure. The uncertainty estimates are based on
  the procedures described in \S~2. We intend for this Table to appear
  in entirety in the electronic version of the journal. A portion is
  shown here to illustrate its format. The data are also available in
  digital form from the authors upon request.}

\end{deluxetable}

\begin{deluxetable}{llll}

\tabletypesize{\small}
\tablecaption{System Parameters of XO-1\label{tbl:params}}
\tablewidth{4.5in}

\tablehead{
\colhead{Parameter} & \colhead{Median value} & \multicolumn{2}{c}{68\% conf.\ limits} \\
\colhead{ }         & \colhead{ }        & \colhead{lower} & \colhead{upper}
}

\startdata
                                     $R_{\rm S}$~[$R_\odot$]& $          0.928$ & $         -0.009$\tablenotemark{a} & $ +          0.015$\tablenotemark{a} \\
                                                            & $               $ & $         -0.013$\tablenotemark{b} & $ +          0.018$\tablenotemark{b} \\
                                                            & $               $ & $         -0.032$\tablenotemark{c} & $ +          0.034$\tablenotemark{c} \\
                                 $R_{\rm P}$~[$R_{\rm Jup}$]& $          1.184$ & $         -0.014$\tablenotemark{a} & $ +          0.025$\tablenotemark{b} \\
                                                            & $               $ & $         -0.018$\tablenotemark{b} & $ +          0.028$\tablenotemark{b} \\
                                                            & $               $ & $         -0.042$\tablenotemark{c} & $ +          0.047$\tablenotemark{c} \\
                                     $R_{\rm P} / R_{\rm S}$& $        0.13102$ & $       -0.00064$ & $ +        0.00064$ \\
                                                         $b$& $           0.14$ & $          -0.10$ & $ +           0.09$ \\
                                                   $i$~[deg]& $          89.31$ & $          -0.53$ & $ +           0.46$ \\
                               $t_{\rm IV} - t_{\rm I}$~[hr]& $          2.992$ & $         -0.015$ & $ +          0.013$ \\
                              $t_{\rm II} - t_{\rm I}$~[min]& $          21.18$ & $          -0.47$ & $ +           0.81$ \\
                                                    $u_1(z)$& $          0.128$ & $         -0.071$ & $ +          0.061$ \\
                                                    $u_2(z)$& $           0.60$ & $          -0.14$ & $ +           0.12$ \\
                                          $2u_1(z) + u_2(z)$& $          0.858$ & $         -0.055$ & $ +          0.049$ \\
                                          $u_1(z) - 2u_2(z)$& $          -1.07$ & $          -0.34$ & $ +           0.30$ \\
                                             $T_c(17)$~[HJD]& $  2453875.92305$ & $       -0.00036$ & $ +        0.00032$ \\
                                             $T_c(18)$~[HJD]& $   2453879.8640$ & $        -0.0011$ & $ +         0.0010$ \\
                                             $T_c(19)$~[HJD]& $  2453883.80565$ & $       -0.00019$ & $ +        0.00017$ \\
                                             $T_c(20)$~[HJD]& $  2453887.74679$ & $       -0.00016$ & $ +        0.00014$
\enddata

\tablecomments{The parameter values in Column 2 are the median values
  of the distributions shown in Fig.~3. The confidence limits in
  Columns 3 and 4 are based on the MCMC analysis.}

\tablenotetext{a}{These uncertainties ignore the uncertainty in
  stellar mass. (For parameters with no designation, the uncertainty
  in the stellar mass is irrelevant.)}

\tablenotetext{b}{These uncertainties include the $0.03~M_\odot$
  uncertainty in the stellar mass reported by \citet{McCullough.2006},
  propagated according to $R_{\rm S}~\propto~(M_{\rm
    S}/M_\odot)^{1/3}$ and $R_{\rm P}~\propto~(M_{\rm
    S}/M_\odot)^{1/3}$.}

\tablenotetext{c}{These uncertainties include a stellar mass
  uncertainty of $0.10~M_\odot$.}

\end{deluxetable}

\end{document}